\documentclass[twocolumn]{aa} 

\usepackage{graphicx}

\usepackage{txfonts}
\begin{document}
\title{Magnetic field amplification in proto-neutron stars}

\subtitle{the role of the neutron-finger instability for dynamo excitation}

\author{L. Naso \inst{1} \and L. Rezzolla \inst{2,3,4} \and A. Bonanno \inst{5} \and L. Patern\`o \inst{6}}

\offprints{L. Naso}

\institute{SISSA - International School for Advanced Studies, Via Beirut 2-4, I-34014 Trieste, Italy 
\and 
Max-Planck Institut f\"ur Gravitationphysik, Albert Einstein Institut, Am M\"uhlenberg 1, D-14476 Golm, Germany
\and
Department of Physics and Astronomy, Louisiana State University, 202 Nicholson Hall, Tower Dr., Baton- Rouge, LA 70803, USA
\and
INFN, Sezione di Trieste, Via A. Valerio 2, I-34127 Trieste, Italy
\and 
INAF -- Astrophysical Observatory of Catania, Via S. Sofia 78, I-95123 Catania, Italy 
\and 
Department of Physics and Astronomy, Astrophysics Section, University of Catania, Via S. Sofia 78, I-95123 Catania, Italy 
}

  \date{Received; accepted}

  \abstract
 {} 
{During the first 40 s after their birth, proto-neutron stars are expected to be subject to at least two types of instability. The first one, the \textit{convective} instability, is excited in the inner regions, where the entropy gradient produces a Rayleigh-type convection. The second one, the \textit{neutron-finger} instability, is instead excited in the outer layers where the lepton gradients are large. Both instabilities involve convective motions and hence can trigger dynamo actions that may be responsible for the large magnetic fields in neutron stars and magnetars. However, because they have rather different mean turbulent velocities, they are also likely to give rise to different types of dynamo.}
{We have solved the mean-field induction equation in a simplified one-dimensional model of both the convective and the neutron-finger instability zones. Although very idealized, the model includes the nonlinearities introduced by the feedback processes that tend to saturate the growth of the magnetic field ($\alpha$-quenching) and suppress its turbulent diffusion ($\eta$-quenching). The possibility of a dynamo action is studied within a dynamical model of turbulent diffusivity where the boundary of the unstable zone is allowed to move. A large number of numerical simulations have been performed in which the relevant parameters, such as the spin-period, the strength of the differential rotation, the intensity of the initial magnetic field, and the extent of the neutron finger instability zone, have been suitably varied.}
{We show that the dynamo action can also be operative within a dynamical model of turbulent diffusivity and that the amplification of the magnetic field can still be very effective. Furthermore, we confirm the existence of a critical spin-period, below which the dynamo is always excited independently of the degree of differential rotation, and whose value is related to the size of the neutron-finger instability zone. We provide a relation for the intensity of the final field as a function of the spin of the star and of its differential rotation.}
{Although they were obtained by using a toy model, we expect that our results are able to capture the qualitative and asymptotic behaviour of a mean-field dynamo action developing in the neutron-finger instability zone. Overall, we find that such a dynamo is very efficient in producing magnetic fields well above equipartition, and thus that it could represent a possible explanation for the large surface magnetic fields observed in neutron stars.}


   \maketitle

%

\section{Introduction \label{INTRODUCTION}}

The present understanding of the processes that produce the large magnetic field strengths observed in neutron stars (NSs) is still far from being complete. Most of the information about their magnetic fields is in fact derived either from their X-ray spectra, or from their spin-down when these NSs are seen as pulsars. While the former reflects a measure of the local surface field $B_{\rm surf}$, the latter provides information on the global dipolar magnetic field $B_{\rm d}$, if the spin-down is assumed to be solely due to dipolar electromagnetic emission. These two measures are not always in agreement, showing that the measured magnetic fields may have very different length-scales and intensities. In particular, they seem to suggest the presence of more intense and small-scale surface magnetic fields, together with less strong globally dipolar ones. For the pulsar 1E 1207.4-5209, for instance, the dipolar magnetic field estimated from the spin-down rate is $B_{\rm d} \sim 2-4\,\times\,10^{12}\,\rm G$ (Pavlov et al. \cite{pavlov02}), while the surface field estimated from the absorption features in its spectrum is $B_{\rm surf} \sim 1.5\,\times\,10^{14}\,\rm G$ (Sanwal et al. \cite{sanwal02}). Similarly, observations of the pulsar RBS B1821-24 (Becker et al. \cite{becker03}) indicate that $B_{\rm d} \sim 10^9\,\rm G$, while $B_{\rm surf} \sim 10^{11}\,\rm G$.

The existence of magnetic fields with different strengths and distributed on different length-scales can be explained in terms of a
dynamo mechanism driven by the simultaneous presence of rotation and turbulent motions. During the first $\sim 40$ s after their birth, proto-neutron stars (PNSs) are expected to develop hydrodynamical instabilities (Epstein \cite{epstein79}; Livio et al. \cite{livio80}; Burrows \& Lattimer \cite{burrows86}), which can excite a hydro-magnetic dynamo. Such instabilities could be essentially of two types. The first one, driven by the entropy gradient, is a Rayleigh-type {\em convective instability} (CI) that operates in the inner regions of the star. The second one is a double diffusive instability, driven by both the entropy and leptonic gradients. This is usually referred to as the {\em neutron-finger instability} (NFI); it operates in the outer regions of the PNS (Miralles et al. \cite{miralles00}; \cite{miralles02}) and is expected to evolve by creating finger-like downflows when the neutrinos are still confined (Mezzacappa et al.\cite{mezzacappa98}). Although some authors have recently raised doubts about the existence of the NFI (Buras et al. \cite{buras06}, Dessart et al. \cite{dessart06}, Bruenn et al. \cite{bruenn07}), no firm conclusion has yet been reached, leaving the debate open. Here, we do not attempt to enter this debate but rather, because of the interesting astrophysical implications that it may have, we will consider the NFI as taking place and having the dynamical properties as described by Miralles et al. (\cite{miralles00}, \cite{miralles02}).

We note that the co-existence of the two instability mechanisms produces both a local dynamo process (Thompson \& Duncan
\cite{thompson93}; Xu \& Busse \cite {xu01}) and a mean-field one (Bonanno et al. \cite{bonanno03}). As shown in Miralles et al. (\cite{miralles00}, \cite{miralles02}), the growth-times of the instabilities in the two regions differ by 2 or 3 orders of magnitudes, being $\tau_{\rm {\mbox{\tiny CI}}} \sim 0.1\,\rm ms$ in the CI zone and $\tau_{\rm {\mbox{\tiny NFI}}} \sim 30-100\,\rm ms$ in the NFI zone.  Since the typical spin period of a PNS is $P \sim 100\,\rm ms$, the turbulent eddies created by the CI are not influenced by the rotation and therefore they can only excite a local dynamo. On the other hand, the Rossby number in the NFI zone, defined as $R_{\rm o}\simeq P/\tau_{\rm {\mbox{\tiny NFI}}}$, is about unity and the turbulent motions can therefore be influenced significantly by the rotation, favoring the excitation of a global mean-field dynamo. Because of the large difference in the growth-rates of the two instabilities, the two processes, i.e. the local dynamo and the global mean-field one, are essentially decoupled.

Here, we focus our attention on the turbulent mean-field dynamo action that may be excited in the NFI zone. More specifically, we exploit a simple one-dimensional (1D) toy model that aims at capturing, at least qualitatively, the features of the dynamo action.  The model, which includes the nonlinearities introduced by the feedback processes, which in turn tend to saturate the growth of the magnetic field, i.e. $\alpha$-quenching (Bonanno et al. \cite{bonanno05}, \cite{bonanno06}; R\"udiger \& Arlt \cite{ruediger96}), and suppress its turbulent diffusion, i.e. $\eta$-quenching (R\"udiger \& Arlt \cite{ruediger96}), is evolved numerically with a very large variety of initial conditions. These include varying the spin period of the PNS, the strength of the differential rotation between the core and the surface, the intensity of the primordial (seed) magnetic field, and the extent of the NFI zone.

Overall, we find that increasing the extent of the neutron-finger instability zone favours the dynamo excitation, and
that the combined action of differential rotation and diffusion can produce an increase in the strength of the generated toroidal field by several orders of magnitude. We also confirm the existence of a critical spin-period, below which the dynamo is always excited independently of the differential rotation strength, and whose value is related only to the size of the neutron-finger instability zone. Since the numerical simulations show that the turbulent mean-field dynamo is very efficient in producing magnetic fields well above the equipartition value, we confirm the results of Bonanno et al. \cite{bonanno06}; even in presence of both a moving boundary of the instability zones and of an $\eta$-quenching, i.e. the proposed mechanism could represent an explanation for the large surface magnetic fields observed in several neutron stars.

The paper is organized as follows: in Sect. \ref{MODEL} we formulate the model and derive the system of partial differential equations, which are then solved with the numerical approach described in Sect. \ref{NMandT}. In Sect. \ref{AandR} we describe the methodology and report numerical results. Finally in the conclusions we summarize our results.

\section{The model \label{MODEL}}

The mean-field induction equation of a turbulent, magnetized and conducting plasma can be written as
\begin{equation}
\partial_t \mathbf{B} = \nabla \times \left( \mathbf{v} \times
\mathbf{B} + \alpha \mathbf{B} \right) - \nabla \times \left(
\eta\nabla\times \mathbf{B} \right)\;,
\label{induction}
\end{equation}
where $\mathbf{B}$ is the mean magnetic field, $\mathbf{v}$ the mean velocity field, $\eta$ the turbulent magnetic diffusivity, and
$\alpha$ a pseudo-scalar measuring the efficiency of the dynamo $\alpha$-effect (this is usually referred to as the $\alpha$-parameter). This effect is related to the generation of a net electromotive force ${\mathcal E}_{turb}$ by the turbulent components of the magnetic field ($\mathbf{B}^\prime$) and the velocity field ($\mathbf{v}^\prime$). More specifically, ${\mathcal E}_{turb}\equiv \overline{ \mathbf{v}^\prime\times \mathbf{B}^\prime }=\alpha_{ij}\bar{B_j}+\beta_{ijk}\partial_k\bar{B_j}+\ldots$  In the case of isotropic turbulence $\alpha_{ij}=\alpha\delta_{ij}$ and $\beta_{ijk}=\beta\epsilon_{ijk}$ and ${\mathcal E}_{turb}$ can be approximated as ${\mathcal E}_{turb} = \alpha\bar{B}-\beta\nabla\times\bar{B}$.

Our simplified model follows closely the one introduced by Brandenburg et al. (\cite{brandenburg89}), R\"udiger et al. (\cite{ruediger94}), and R\"udiger \& Arlt (\cite{ruediger96}). It uses orthogonal Cartesian coordinates, and the PNS is modeled has having planar symmetry with the $z$-axis being the axis of rotation, the $(x,y)$-plane representing the stellar equatorial plane, and $H$ the semi-height being the radius of the PNS. 

\begin{figure}[t]
\begin{center}
\resizebox{\hsize}{!}{\includegraphics{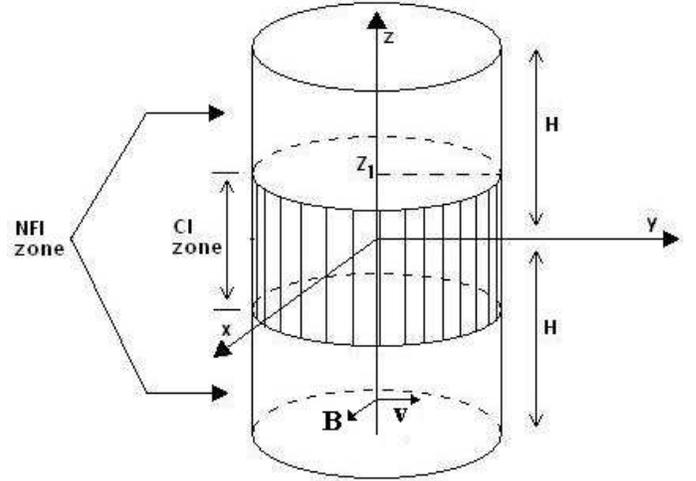}}
\caption{Schematic diagram of our model of the PNS.}
\label{model}
\end{center}
\end{figure}

Clearly, this simplified model has the advantage of leading to a very simple expression for the induction equation (\ref{induction}) which, after making a suitable choice for the velocity and magnetic fields, can be recast into a system of two coupled partial differential equations that are of first order in time. This reduces the computational costs enormously and allows a parametric investigation to be performed, which would be impossible if it involved fully 3D simulations. Yet, despite the considerable simplifications, this simple geometry has been shown to yield instructive results that have opened the way for modern, realistic 3D models (Weiss et al. \cite{weiss84}). Our simplified model of the PNS is shown in Fig. \ref{model}, where we highlight the CI and NFI zones that are separated by a thin interface of thickness $d$, which is not shown in the figure. Note that despite the cylindrical appearance of the star, the system does not have a cylindrical symmetry but rather a planar one across the $(x,y)$ plane.

Another important aspect in the present modeling of the turbulent mean-field dynamo is the so-called kinematic dynamo
approximation, in which the solution of the induction Eq. (\ref{induction}) is assumed to be decoupled from the Euler equations
(i.e. the PNS is assumed to be in hydrostatic equilibrium at all times), and no feedback from the magnetic field is taken into account in the conservation of momentum and energy. As a result, the velocity field is taken as pre-assigned and time-independent. This approximation works quite well as long as the magnetic field strength is small (i.e. as long as the magnetic pressure and tension are negligible in the Euler equations), but has the drawback that nothing prevents the indefinite growth of the magnetic field once a dynamo action is present. In real systems, the velocity profile will adjust itself in such a way as to reduce the efficiency of the dynamo and here, as a way of mimiking this feedback, we introduce two \textit{``quenching functions''} (described later in more detail) that suppress the amplification as the total field increases above a certain threshold, thus leading the system towards saturation even in the absence of a consistent feedback.

Our model for the kinematic dynamo follows the one proposed by R\"udiger et al. (\cite{ruediger94}) and Blackman \& Brandenburg
(\cite{blackman02}) and, in particular, we consider a velocity field 
\begin{displaymath}
\mathbf{v}=\left( 0,kx,0 \right) \;,
\end{displaymath}
and a magnetic field having components only in the $(x,y)$ plane but a with a vertical dependence
\begin{displaymath}
\mathbf{B}=\left( B_x(z),B_y(z),0 \right) \;.
\end{displaymath}

\begin{figure}[t]
\begin{center}
\resizebox{\hsize}{!}{\includegraphics{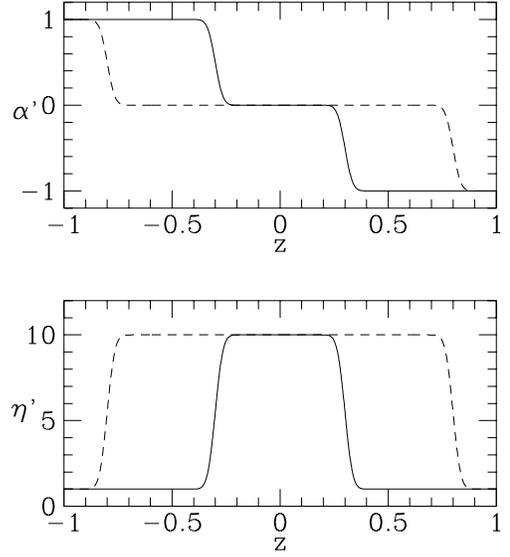}}
\caption{Top panel: profiles of the normalized function $\alpha^\prime$ at the beginning of the numerical simulation for $t=0$
(solid line) and at the end of the numerical simulation for $t=40\,\rm s$ (dashed line); bottom panel: the same as in the top panel but for $\eta^\prime$.}
\label{aeprofiles}
\end{center}
\end{figure}

The $\alpha$-parameter and the magnetic diffusivity $\eta$ are both expressed as the product of three terms
\begin{eqnarray}
\alpha &=& \alpha_0 \; {\alpha^\prime}(z,t) \;
	\psi_{\alpha}(\mathbf{B})\;,
\label{alphaeq} \\
\eta &=& \eta_0 \; {\eta^\prime}(z,t) \; \psi_{\eta}(\mathbf{B})\;,
\label{etaeq}
\end{eqnarray}
where $\alpha_0$ and $\eta_0$ measure the strength of the $\alpha$-effect and turbulent diffusion respectively, while
${\alpha^\prime}(z,t)$ and ${\eta^\prime}(z,t)$ represent the profiles of $\alpha$ and $\eta$ in the two instability regions. More
specifically, ${\alpha^\prime}$ is chosen to be antisymmetric across the equatorial plane and different from zero only in the NFI zone, where the mean-field dynamo is at work. The turbulent diffusivity ${\eta^\prime}$, on the other hand, is set to be of the order of unity in the NFI zone and about an order of magnitude larger in the turbulent CI zone. We implement these prescriptions by making use of the error function ``erf'' as
\begin{eqnarray}
{\alpha^\prime}(z,t)\equiv
\left\{
\begin{array}{ll}
\frac{1}{2}\left[1+{\rm erf}\left(-(z+\lambda)/d\right) \right]\;, \qquad
	&\mbox{ $z \in [-H,0]$} \\ 
&\\
-\frac{1}{2}\left[1+{\rm
  	erf}\left((z-\lambda)/d\right) \right]\;, \qquad
	&\mbox{ $z \in [0,H]$}
\end{array} 
\right.
\label{alphaprime}
\end{eqnarray}
\begin{eqnarray}
{\eta^\prime}(z,t)\equiv
	\frac{1}{10}\left\{10-\frac{9}{2}\Big[1+{\rm
    erf}\left[(z-\lambda)/d\right]\Big]\right\} \; \times
\label{etaprime}
\nonumber \\ 
\left\{10-\frac{9}{2}\Big[1+{\rm erf}
	\left[-(z+\lambda)/d\right]\Big]\right\}\;,
\end{eqnarray}
where $\lambda \equiv z_1+Vt$, with $z_1$ being the coordinate of the boundary between the CI and NFI zones and $V$ the expansion velocity of the boundary layer. The quantity $d$ represents the thickness of the interface between the CI and NFI zones and is used to obtain a smooth change of the error function, with smaller values leading to sharper changes; for the results reported here we have chosen $d/H=0.04$. The profiles of $\alpha^\prime$ and $\eta^\prime$, as given by Eqs. (\ref{alphaprime}) and (\ref {etaprime}), are shown in Fig. \ref{aeprofiles}, with the solid and dashed lines indicating the initial conditions and the final conditions after $40\,\rm s$, respectively.

Finally, $\psi_\alpha(\mathbf{B})$ and $\psi_\eta(\mathbf{B})$ appearing in Eqs.~(\ref{alphaeq}) and (\ref{etaeq}) represent the
quenching functions for the $\alpha$-effect and turbulent diffusion $\eta$, respectively. These terms are used to limit the otherwise
unlimited growth of the magnetic field ($\alpha$-quenching) and suppress its turbulent diffusion ($\eta$-quenching). In general they
are expected to have a different dependence on the magnetic field strength, but we here consider a single expression for the two
functions in terms of the equipartition magnetic field $B_{\rm eq}=\langle {\rm u} \rangle \sqrt{4\pi\rho}$, where $\langle \rm u
\rangle$ is the mean velocity of turbulent eddies and $\rho$ the mass-density
\begin{equation}
\psi_{\alpha,\eta}(\mathbf{B})\equiv
	\left[1+\int_{-H}^H
	\left(\frac{B}{B_{\rm eq}}\right)^2 dz\right]^{-1} \;.
\end{equation}

Making use of these definitions, Eq.~(\ref{induction}) can be split into the following two coupled scalar partial differential equations
\begin{eqnarray}
\partial_t B_x &=& -\partial_z (\alpha B_y) + \partial_z\left(\eta
	\partial_zB_x\right) \;,
\label{DIM1}\\
\partial_t B_y &=& \partial_z(\alpha B_x)+B_x\partial_x v_y
	+\partial_z\left(\eta\partial_zB_y\right) \;,
\label{DIM2}
\end{eqnarray}
which can also be written in a dimensionless form by scaling lengths in units of the semi-height of the cylinder $H$, times in units of the diffusion time $\tau_{_{\rm D}}=H^2/\eta_0$, and magnetic fields in units of $B_{\rm eq}$. We also find it useful to introduce
the dimensionless parameters 

\begin{equation}
C_\alpha \equiv \alpha_0 \frac{H}{\eta_0}\;, 
	\qquad {\rm and} \qquad
C_\Omega \equiv \partial_x v_y \left(\frac{H^2}{\eta_0}\right) = 
	k\left(\frac{H^2}{\eta_0}\right)\;, 
\end{equation}
which represent the Reynolds numbers for the $\alpha$-effect and the differential rotation, respectively. Furthermore, by introducing the standard vector potential $\mathbf{A}=[A_x(z),A_y(z),0]$, so that the poloidal component of the magnetic field is ${\cal B}_p \equiv B_x = -\partial_z A_y$, we obtain the following dimensionless equations for the vector potential $\mathcal{A}=A_y$ and the toroidal component of the magnetic field $\mathcal{B}_t \equiv B_y$
\begin{eqnarray}
\partial_t \mathcal{A} =
C_\alpha\alpha(z,t)\psi_{\alpha}(B_{tot})\mathcal{B}_t+\eta(z,t)
\psi_{\eta}(B_{tot})\partial_z^2\mathcal{A}
\label{EQFINA} \\
\partial_t \mathcal{B}_t = -
C_\alpha\partial_z\left[\alpha(z,t)\psi_{\alpha}(B_{tot})\partial_z
  \mathcal{A}\right] - C_\Omega\partial_z \mathcal{A} + \nonumber
\\ +\; \partial_z\left[\eta(z,t)\psi_{\eta}(B_{tot})\partial_z
  \mathcal{B}_t \right]
\label{EQFINB}
\end{eqnarray}
where $B_{tot} \equiv [\mathcal{B}_t^2 + (\partial_z \mathcal{A})^2]^{1/2}$. Once the initial conditions $\mathcal{B}_t(z,0)$ and $\mathcal{A}(z,0)$ are given, together with the parameters, $C_\alpha$, $C_\Omega$, $\alpha$, $\eta$, $\psi_\alpha$, $\psi_\eta$, and suitable boundary conditions at the stellar edges, it is possible to solve Eqs. (\ref{EQFINA}) and (\ref{EQFINB}) to describe the time evolution of the magnetic field. Our choice for the boundary conditions reflects the fact that we are interested in an adiabatic evolution of the magnetic field in the stellar interior, thus neglecting the energy losses related to a Poynting flux. Because of this, we simply set the magnetic fields at these locations to zero.

It is important to note that the parameters $C_\alpha$ and $C_\Omega$ are not linearly independent but can be related in terms of the strength of the differential rotation, of the pressure scale-height and of the ``radius'' of the PNS. To deduce this relation
we recall that the $\alpha\Omega$-dynamo assumes that $\alpha_0\sim \Omega L_p$, where $\Omega$ is the angular velocity and $L_{\rm p}$ the pressure scale-height, so that 
\begin{equation}
P \sim \frac{\xi}{C_\alpha} = \frac{2\pi L_{\rm p}H}{C_\alpha\eta_0}\;,
\label{PCalpha}
\end{equation}
where $\xi \equiv 2\pi L_{\rm p}H/\eta_0$. For a typical PNS with mass $\sim 1 M_{\odot}$, $H\sim 15\,{\rm km}$ and $L_{\rm p} \sim 3\,{\rm km}$ (Bonanno et al. \cite{bonanno03}), $\eta_0 \sim L_{\rm p}^2/\tau_{\rm {\mbox{\tiny NFI}}}$ ranges from $9\times 10^{11}\,\rm cm^2s^{-1}$ for NFI eddy turnover times of $100\,\rm ms$, up to $3\times 10^{12}\,\rm cm^2s^{-1}$ for NFI eddy turnover times of $30\,\rm ms$. Such values of $\eta_0$ yield typical diffusion timescales $\tau_{_{\rm D}}$ ranging from $0.75\,\rm s$ to
$2.5\,\rm s$, respectively. As a result, the parameter $\xi$ is expected to be roughly in the range $1\,\rm s \leq \xi \leq 3\,\rm s$
for all of the relevant parameter space considered here.

In a similar way, since $\partial_x v_y = k \sim \Delta\Omega$, and defining the relative differential rotation strength as $q \equiv
\Delta\Omega/\Omega \simeq (\Omega_{\rm s}-\Omega_{\rm c})/\Omega_{\rm s}$, where $\Omega_{\rm s}$ and $\Omega_{\rm c}$ are the angular velocities of the surface and the core respectively, it is possible to conclude that
\begin{equation}
q \sim  \zeta \frac{C_\Omega}{C_\alpha} \;, 
\label{qfactor}
\end{equation}
where $\zeta \equiv L_{\rm p}/H \simeq 1/5$. Here we use $\Delta\Omega$ as a global measure of the rotational stress instead of using a necessarily arbitrary function that describes the behaviour of $\Omega(z)$ in the region between $\Omega_{\rm c}$ and $\Omega_{\rm s}$. On the other hand we are severely limited by our ignorance of the detailed processes leading to the appearence of the differential rotation in the NFI zone of PNSs.

For all of the calculations reported here we assume that the angular velocity of the core is larger than that of the surface, so that $q < 0$ and, conservatively, we limit our analysis to values of $|q|$ not exceeding $10^2$, i.e. a core rotating $10^2$ times faster than surface. Finally, by assuming a mass-density in the NFI zone $\rho \sim 10^{13}\,\rm g\,cm^{-3}$ and that the eddy convective velocities $\langle {\rm u} \rangle \simeq L_{\rm p}/\tau_{\rm {\mbox{\tiny NFI}}} \sim 3\times 10^{6}\,\rm cm\,s^{-1}$, it follows that $B_{\rm eq}$ is of the order of $10^{13}\,{\rm G}$. The parameters defined in Eqs. (\ref{PCalpha}) and (\ref{qfactor}) essentially determine the parameter space for the solutions of Eqs. (\ref{EQFINA}) and (\ref{EQFINB}).

\section{Numerical method and tests\label{NMandT}}

In order to solve the mixed parabolic-hyperbolic system of partial
differential equations~(\ref{EQFINA}) and (\ref{EQFINB}), we
discretize the continuum space-time by replacing it with a two
dimensional grid, where the two dimensions represent the space and the
time variables, $z$ and $t$, respectively. We use constant spacing in
both directions, with a typical grid of 50 zones. Tests were performed with a larger number
of gridpoints (100, 200 and 400) and have revealed that a minimum of 50
gridpoints was sufficient to yield a small-enough truncation error.
The evolution algorithm chosen is the FTCS (Forward-in-Time,
Centered-in-Space) scheme, which gives a first-order approximation for
the time derivatives and a second-order approximation for the space
derivatives. Furthermore, stability requires the timestep to be
$\Delta t = {\cal O}(\Delta z^2)$ (we typically use $\Delta t =
10^{-2}\Delta z^2$), thus making the whole algorithm second-order both
in space and in time. Using ghost-zones for implementing the boundary
conditions, the final form of the finite-difference equations is
\begin{eqnarray}
\mathcal{A}_{j}^{n+1}&=&\mathcal{A}_{j}^{n}+\Delta t\,d_j^n\,\mathcal{B}_{j}^{n} + \nonumber\\
&&+ k_1\,a_j^n\,\left( \mathcal{A}_{j+1}^{n}-2\,\mathcal{A}_{j}^{n}+\mathcal{A}_{j-1}^{n}\right)\;,\\
\mathcal{B}_{j}^{n+1}&=&\mathcal{B}_{j}^{n} +k_2\,c\,\left( \chi_{j+1}^n -\chi_{j-1}^n \right) +\nonumber\\
&&+\; k_2\,b\,\left( \mathcal{A}_{j+1}^{n}-\mathcal{A}_{j-1}^{n} \right)+\nonumber\\
&&+\; k_1\,a_j^n\,\left( \mathcal{B}_{j+1}^{n}-2\,\mathcal{B}_{j}^{n}+\mathcal{B}_{j-1}^{n}  \right)+\nonumber\\
&&+\; k_2\,f_j^n\,\left( \mathcal{B}_{j+1}^{n}-\mathcal{B}_{j-1}^{n} \right)\;,
\end{eqnarray}
where
\begin{eqnarray}
\chi_j^n=e_j^n\,\frac{\mathcal{A}_{j+1}^{n}-\mathcal{A}_{j-1}^{n}}{2\Delta z}
\end{eqnarray}
and
\begin{eqnarray}
&&k_1=\frac{\Delta t}{\Delta z^2}\mbox{,}\quad k_2=\frac{\Delta t}{2\,\Delta z}\\
&&a_j^n=\eta_j^n\,\psi_\eta^n \mbox{,}\quad b=-C_\Omega \mbox{,}\quad c=-C_\alpha\\
&&d_j^n=C_\alpha\,\alpha_j^n\,\psi_\alpha^n \mbox{,}\quad e_j^n=\alpha_i^n\,\psi_\alpha^n \mbox{,}\quad f_j^n=\frac{\eta_{j+1}^n-\eta_{j-1}^n}{2\Delta z}
\end{eqnarray}

As a test of the code we have considered the equations when $C_\alpha=C_\Omega=0$, $\psi_\eta=1$ and $\eta(z,t)=5 \cdot 10^{-2}$ so as to have two decoupled purely parabolic equations, and compared the numerical solution with the analytic one. The result of the comparison is that the maximum error is of the order of $2\%$. 

We have also checked the convergence of the method by comparing the numerical solutions obtained with different spatial grids at a given time. The same continuum function $u(z,t)$ is approximated in a different way according to the space-time discretization: $u(z,t)=u_i^n(h)+\epsilon_h$, where $\epsilon_h$ is the truncation error. If we suppose that this error depends on the chosen space-interval $h$ only, we can write: $\epsilon_h=k\,h^p$, with $p$ being the order of convergence. It is then possible to deduce the following relation among the values of different discretized functions calculated at the same grid location $i$:
\begin{equation}
\frac{u_i^n(h)-u_i^n(h/2)}{u_i^n(h/2)-u_i^n(h/4)}=2^p\;.
\end{equation}
The value of $p$ obtained as averaged over the spatial domain is $2.00
\pm 0.01$, thus demonstrating a second-order convergence.

\section{Analysis and results \label{AandR}}

While this analysis aims at a better understanding of the behaviour of the dynamos operating in the first stages of the life of a PNS, it is a long way from reproducing realistic conditions. This is partly due to the simplicity of the model employed, and partly to the still poorly constrained physical conditions of a newly born PNS. We recall that, according to Miralles et al. (\cite{miralles00}, \cite{miralles02}), the NFI is expected to last only about $40\,\rm s$, during which the NFI zone goes from occupying a large fraction of the envelope, to being confined to a small layer and then disappearing completely. We model this by assuming that the initial position of the NFI-CI boundary layer is at $z_1^i = 0.3H$ and the final one, after $40\,\rm s$, is at $z_1^f = 0.8H$ (\textit{cf.} Fig.~\ref{model}), with an average expansion velocity of the layer that is $V=(z_1^f-z_1^i)/t= 188\,\rm m\,s^{-1}$ for $H=15\,\rm km$. In this case we also find it convenient to express all variables in terms of dimenionless quantities after introducing $z^\prime \equiv z/H$, $t^\prime \equiv t/\tau_{_{\rm D}}$, and $V^\prime \equiv V\tau_{\rm {\mbox{\tiny D}}}/H$. As a result, the coordinate position of the CI-NFI layer appearing in Eqs. (\ref{alphaprime}) and (\ref {etaprime}) can be written as $\lambda = z_1^\prime + V^\prime t^\prime = 0.3 + 0.0125\,({\rm s}^{-1})\,t\,({\rm s})$, where $-1\leq z^\prime \leq 1$,
and $0\leq t \leq 40\,\rm s$, corresponding to a number of diffusion times ranging from $16$ to $53$, depending on the turnover time of the NFI eddies.

\subsection{Initial models\label{IMs}}

As a representative sample of initial data we have considered six different models, four of which have the size of the NFI zone being
constant in time, with either $z_1^\prime=0.3$ (large instability zone) or $z_1^\prime=0.8$ (small instability zone), and which are therefore referred to as \textit{static}. For each of the two values of $z^\prime_1$ we have examined the behaviour with and without $\eta$-quenching. Besides the static configurations, which are useful for studying the extreme cases of thick and thin NFI zones respectively, we have also considered two cases, which are referred to as \textit{dynamical}, in which the NFI is allowed to shrink in time from the initial value of $z_1^\prime=0.3$ to the final one of $z_1^\prime=0.8$, over the $40\,\rm s$ during which the instability is expected to be active. For these dynamical models we have also studied the effect of activating or not activating the $\eta$-quenching. Such cases are useful for studying the role played by a dynamically shrinking NFI zone in the onset of the dynamo action and in the final magnetic fields obtained.

The static models are indicated as A, B, Aq, Bq, while the dynamical ones are indicated as AB and ABq; in all cases, the letter ``q'' is used to indicate whether or not $\eta$-quenching is taken into account. For all of these models we have carried out a large number of simulations by varying the seed magnetic field $B_{\rm s}$, as well as the spin period $P$ and the strength of the differential rotation $|q|$; all the models have been evolved for $40$ diffusion times ({i.e. between $30 \, {\rm s}$ and $100 \, {\rm s}$). A summary of the properties of the different initial models and of the parameters used in the simulations is presented in Table \ref{configurations}.

\begin{table}
\caption{The parameters $z_1^\prime$ (initial position of the boundary layer), $V^\prime$ (mean velocity of the boundary layer), and $\eta$-q ($\eta$-quenching activated or not) that define the configurations analyzed, and the ranges of values of $B_{\rm s}$, $C_\alpha$ and $|q|$ used for the simulations. As regards $C_\alpha$, we mainly used values between $5$ and $200$, since for smaller values the dynamo is not excited and for larger values the spin period would be too short.}
\begin{center}
\begin{tabular}{ccccccc}
\hline
\hline
    & $z_1^\prime$ & $V^\prime$ & $\eta$-q & $B_{\rm s}/B_{\rm eq}$ & $C_\alpha$ & $|q|$\\ 
\hline
A   & $0.3$ & $0.0$ & no  & $10^{-7}-10^{-1}$ & $\leq 10^3$ & $10^{-8}-10^2$\\
B   & $0.8$ & $0.0$ & no  & $10^{-7}-10^{-1}$ & $\leq 10^3$ & $10^{-8}-10^2$\\
Aq  & $0.3$ & $0.0$ & yes & $10^{-7}-10^{-1}$ & $\leq 10^3$ & $10^{-8}-10^2$\\
Bq  & $0.8$ & $0.0$ & yes & $10^{-7}-10^{-1}$ & $\leq 10^3$ & $10^{-8}-10^2$\\
\hline
AB  & $0.3$ & $0.0125$ & no  & $10^{-7}$ & $\leq 10^3$ & $10^{-8}-10^2$\\
ABq & $0.3$ & $0.0125$ & yes & $10^{-7}$ & $\leq 10^3$ & $10^{-8}-10^2$\\
\hline
\end{tabular}
\end{center}
\label{configurations}
\end{table}

\subsection{Time evolution and critical period \label{TEandCP}}

The time evolutions of the average toroidal field ${\cal B}_t$ and poloidal field ${\cal B}_p$ (calculated in terms of their 2-norms) are shown in Figs. \ref{normB} and \ref{normdA} respectively for the configuration Aq, with $|q|=2$ and $C_\alpha =4$ ($0.25\,\rm s\leq P\leq 0.75\,\rm s$), and for different values of the seed magnetic field $B_{\rm s}$ in the range $10^{-7}B_{\rm eq}\leq B_{\rm s}\leq 10^{-1}B_{\rm eq}$.

For all of the configurations examined, the evolution of the magnetic field is rather similar and can be separated into two main stages: the first one is a transient phase during which the dynamo action amplifies the seed magnetic field exponentially; in the second phase the magnetic field instead reaches saturation around the equipartition value through the back reaction of the $\alpha$-quenching. In general, the field reaches saturation within the $40\,\rm s$ lifetime of the NFI zone, except when the seed magnetic field is lower than $10^{-7}B_{\rm eq}$ and $|q| \leq 2$. Nevertheless, in these cases the final magnetic value is of the order of $10^{-2}\,B_{\rm eq}$, thus corresponding to $10^{11}\,{\rm G}$.

These two stages can easily be distinguished in Figs. \ref{normB} and \ref{normdA}, which report the time-evolution of the toroidal and poloidal magnetic fields, respectively. The secular slope of the curves in the exponential-amplification phase is clearly independent of $B_{\rm s}$ and constant in time, with $B=B_{\rm s}\;{\rm e}^{~t/\tau_{amp}}$ and $\tau_{amp}\sim0.6\,\rm \tau_D$, where $\tau_D \simeq 25 \tau_{\rm NFI}$ is the diffusion timescale. The growth-time however depends on $|q|$ and $C_\alpha$, thus suggesting that the seed magnetic field determines only the time interval necessary to achieve the saturation, but not the final strength of the magnetic field. Clearly, if the initial field is too small, the dynamo cannot reach the saturation phase within $40 \, {\rm s}$. Other values of the growth times are $\tau_{amp}\sim0.98\,\rm \tau_D$ for configuration ABq (with $C_\alpha=4$ and $|q|=3$) and $\tau_{amp}\sim2.19\,\rm \tau_D$ for configuration B (with $C_\alpha=29$ and $|q|=4$).

\begin{figure}[!ht]
\begin{center}
\resizebox{\hsize}{!}{\includegraphics{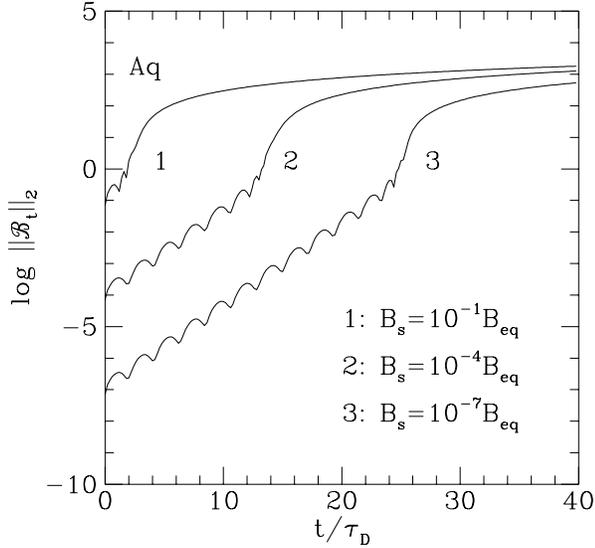}}
\caption{Behaviour of the 2-norm of ${\cal B}_t$ as a function of the number of diffusion times ($t/\tau_D$) for the Aq configuration, $|q| = 2$, $C_\alpha=4$ ($250\,{\rm ms} \leq P \leq 750\,{\rm ms}$), and for different seed magnetic field strengths, $B_{\rm s}$. The growth-rate of the magnetic field during its amplification phase, before reaching the saturation, is almost independent of the seed magnetic field.}
\label{normB}
\end{center}
\end{figure}

\begin{figure}[!ht]
\begin{center}
\resizebox{\hsize}{!}{\includegraphics{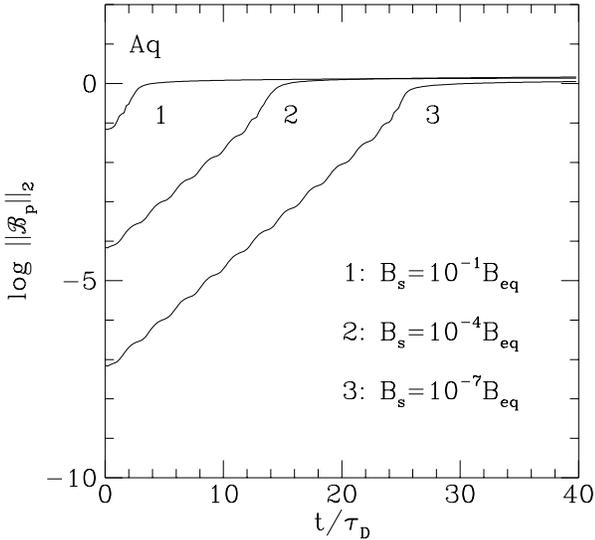}}
\caption{The same as in the Fig. \ref{normB} but for ${\cal B}_p$.}
\label{normdA}
\end{center}
\end{figure}

It is possible that the combination of the rotation rate and differential rotation is not adequate to excite the dynamo and that the magnetic field is not amplified but rather decays with time. For each value of $|q|$ it is possible to find a threshold value of $C_\alpha$ such that for higher rotation rates (shorter periods) the dynamo is excited, while for lower rotation rates (longer periods) it is not. This value of $C_\alpha$ defines a \textit{critical period} $\widetilde{P}_{c}=P_c(|q|)$ through $\xi$ [see Eq. (\ref{PCalpha})].

The presence of a critical period within the space of parameters is shown in Fig. \ref{period_c} for the configuration A. The
behaviour of the dynamo solutions is plotted as a function of the PNS spin period $P$ and differential rotation strength $|q|$ for three different values of $\xi$ (corresponding to different turnover times of the convective eddies in the NFI zone: $\xi=1$, $\tau_{\rm NFI}\simeq 30\,\rm ms$; $\xi = 2$, $\tau_{\rm NFI}\simeq 65\,\rm ms$ and $\xi=3$, $\tau_{\rm NFI}\simeq 100\,\rm ms$). The three curves represent the thresholds between the regions in which the solutions of the dynamo equations grow in time (regions below the curves), thus allowing dynamo excitation, and the regions in which the solutions decay, thus preventing dynamo excitation. The three lines in Fig. \ref{period_c} define therefore the critical dynamo period as a function of $|q|$, and it is evident that the critical spin velocity above which the dynamo operates decreases with increasing differential rotation. The behaviour of these curves suggests that is possible to define a \textit{global} critical period $P_c$ as the minimum of $\widetilde{P}_{c}$, so that if a PNS is rotating with a period shorter than $P_c$ then for that PNS the dynamo will be excited.

The values of the global critical period obtained for the four static configurations are summarized in Table \ref{Pc}. Note that for a fixed value of $\xi$ these periods depend only on the thickness of the NFI zone and are thus independent of the presence of $\eta$-quenching.  As a result, in PNSs with larger NFI zones, the dynamo action will be excited more easily, and thus at lower rotation rates, than for PNSs having smaller NFI zones.

\begin{table}
\caption{For each {\em static} configuration the critical value of $C_\alpha$ ($C_{\alpha}^C$) and the corresponding global critical
period of the PNS, $P_c=\xi/C_{\alpha}^C\,\rm ms$, for $\xi\in[\rm 1,3]$ are reported.}
\begin{center}
\begin{tabular}{ccc}
\hline
\hline
    &  $C_{\alpha}^C$ & $P_c\,[\rm ms]$ \\
\hline
A   &  $5$     				& $200-600$  \\
B   &  $30$   				& $33-99$    \\
Aq  &  $5$     				& $200-600$  \\
Bq  &  $30$   				& $33-99$    \\
\hline
\end{tabular}
\end{center}
\label{Pc}
\end{table}

\begin{figure}[!ht]
\begin{center}
\resizebox{\hsize}{!}{\includegraphics{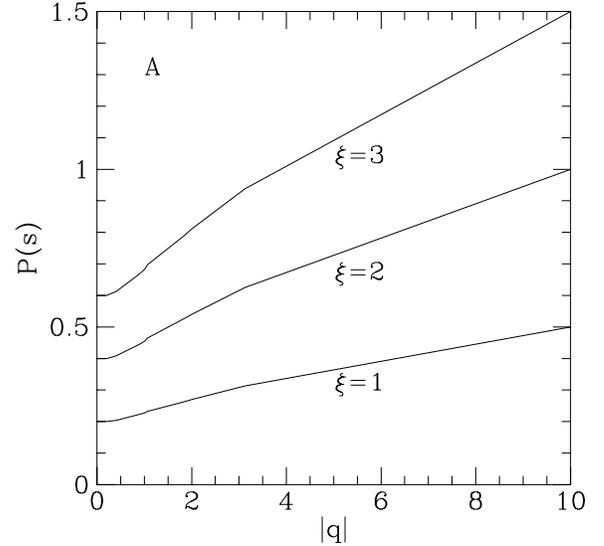}}
\caption{Behaviour of the dynamo solutions as a function of the PNS spin period $P$ and of the differential rotation strength $|q|$ for three different values of $\xi$, related to the turnover time of the convective eddies in the NFI zone.}
\label{period_c}
\end{center}
\end{figure}

\subsection{The role of thickness of the NFI zone \label{NFI}}

Since the NFI zone is the region where the mean field dynamo is at work, it is natural to expect that by increasing the size of this part of the star the efficiency of the dynamo will also increase. To quantify this ``improved efficiency'' we can compare the results obtained for two configurations: one in which the NFI zone occupies $70\%$ of the star (configuration A) and another one in which it covers $20\%$ (configuration B). Reducing the extent of the active part of the star has two main effects: reducing the critical period and reducing the intensity of the final magnetic field.

The first of these effects has already been discussed in the previous section, and from Table \ref{Pc} it is possible to note that reducing the NFI zone by a factor of $3.5$ decreases the critical period by a factor of $6$. To quantify the second effect we consider the ratio between the final strength of the magnetic field in configurations A and B, as a function of the differential rotation parameter $|q|$ and of $C_\alpha$. As expected, the toroidal magnetic field for the configuration A is larger than the one for configuration B, regardless of any other parameter. The value of the ratio is reported in Table \ref{NFIW} and it should be noted that it is really only the toroidal-field ratio that changes with $C_\alpha$, with the poloidal-field ratio being essentially independent of the differential rotation or the spin period.

\begin{table}
\caption{Ratio of the final intensity of the magnetic field in configurations A and B. The values in the table are calculated for
$|q|>10$. For $|q|<10$ we still have ${\cal B}_t^{\rm fin}(A)/{\cal B}_t^{\rm fin}(B)>1$ and ${\cal B}_p^{\rm fin}(A)/{\cal B}_p^{\rm
fin}(B)<1$, but the exact value depends on the differential rotation.}
\begin{center}
\begin{tabular}{cccc}
\hline
\hline
  & \multicolumn{3}{c}{$C_{\alpha}$ } \\
  \cline{2-4}
  &  $50$ & $100$ & $200$ \\
\hline
${\cal B}_t^{\rm fin}(A)/{\cal B}_t^{\rm fin}(B)$   &  $13.4$ & $6.6$  & $13.0$ \\
${\cal B}_p^{\rm fin}(A)/{\cal B}_p^{\rm fin}(B)$   &  $0.38$ & $0.37$ & $0.37$ \\
\hline
\end{tabular}
\end{center}
\label{NFIW}
\end{table}

\subsection{Asymptotic states of the static configurations \label{Static}}

The value of the magnetic field after $40\,\rm s$ of evolution for some representative cases of the static configurations is shown in
Figs. \ref{A_tor}--\ref{A_Aq_pol}, which report the final intensity of the toroidal and poloidal components as functions of $|q|$ and for different values of $C_\alpha$.

Figures~\ref{A_tor} and \ref{A_pol} refer to configurations without $\eta$-quenching and show that the
qualitative evolution of the field is independent of the rotation rate of the star. Furthermore, for small values of the differential
rotation, i.e., for $|q| \lesssim |q^*|$, where $|q^*|$ is a representative threshold value, both the toroidal ${\cal B}_t$ and poloidal ${\cal B}_p$ components of the magnetic field are constant with $|q|$, while for $|q| \gtrsim |q^*|$,  ${\cal B}_t$ begins to increase and ${\cal B}_p$ to decrease, following power-laws with exponents of the same magnitude but opposite signs (see Table \ref{slopes}). Note that the value of $|q^*|$ is smaller for high spin rates since the dynamo is more easily excited when the star is rapidly rotating.

\begin{figure}[!ht]
\begin{center}
\resizebox{\hsize}{!}{\includegraphics{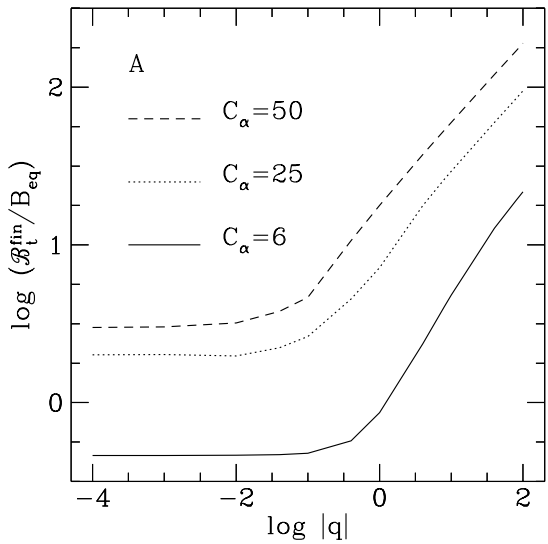}}
\caption{Intensity of the final toroidal field ${\cal B}_t$ in units of $B_{\rm eq}$, at the end of the $40\,\rm s$ NFI evolution time, as a function of the differential rotation parameter $|q|$ for different values of $C_\alpha$. The simulation refers to the configuration A, thus without $\eta$-quenching.}
\label{A_tor}
\end{center}
\end{figure}

\begin{figure}[!ht]
\begin{center}
\resizebox{\hsize}{!}{\includegraphics{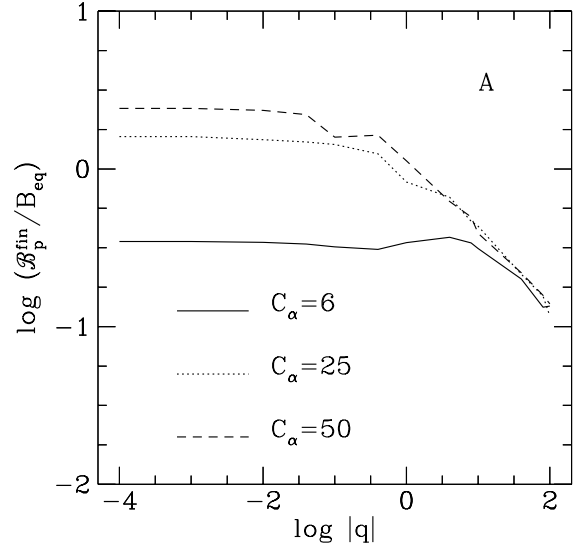}}
\caption{The same as in Fig. \ref{A_tor} but for ${\cal B}_p$.}
\label{A_pol}
\end{center}
\end{figure}

Figures~\ref{A_Aq_tor}--\ref{A_Aq_pol}, on the other hand, show the influence of $\eta$-quenching on the mean-field dynamo by
comparing the final intensities of the magnetic fields for configurations for which the quenching is either active or not, i.e., configurations Aq and A, respectively. It is quite evident that the main consequence of $\eta$-quenching is to increase the amplification factor of the dynamo by several orders of magnitude, leaving the qualitative behaviour of the magnetic field unchanged, except for a value of $|q^*|$ significantly smaller.

\begin{figure}[!ht]
\begin{center}
\resizebox{\hsize}{!}{\includegraphics{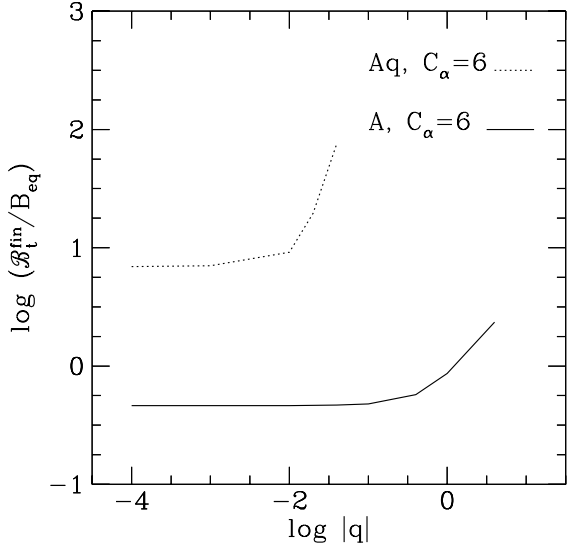}}
\caption{Intensity of the toroidal field ${\cal B}_t$ in units of
$B_{\rm eq}$, at the end of the $40\,\rm s$ NFI evolution time,
plotted as a function of the differential rotation parameter $|q|$ for
$C_\alpha =6$ ($167\,{\rm ms}<P<500\,{\rm ms}$), and an initial seed
magnetic field of $10^{-7}\,B_{\rm eq}$. The simulation shows the
comparison between a configuration with $\eta$-quenching (Aq) and
without $\eta$-quenching (A).}
\label{A_Aq_tor}
\end{center}
\end{figure}

\begin{figure}[!ht]
\begin{center}
\resizebox{\hsize}{!}{\includegraphics{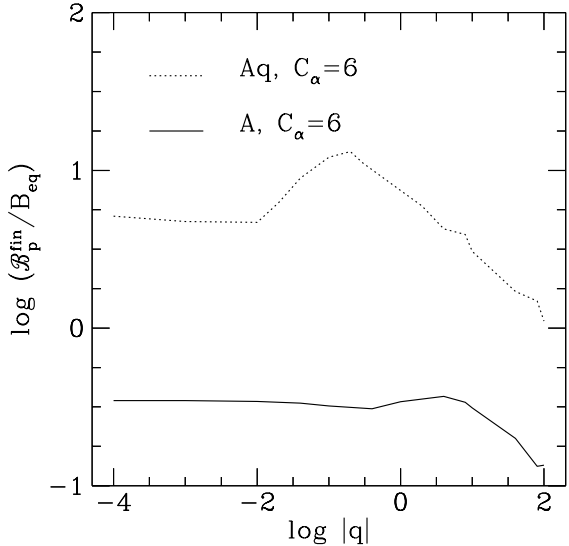}}
\caption{The same as in the Fig. \ref{A_Aq_tor} but for ${\cal B}_p$.}
\label{A_Aq_pol}
\end{center}
\end{figure}

A way of interpreting these results is to recall that the inclusion of $\eta$-quenching enhances the transformation of the poloidal magnetic field into a toroidal one (i.e., $\Omega$-effect) and therefore enhances the importance of differential rotation. In addition, by reducing the magnetic diffusion, $\eta$-quenching effectively favors the amplification of the magnetic field (which becomes essentially frozen with the fluid), increasing the exponential growth-rate of the intensity; in the case shown in Fig. \ref{norm_dA_AAq} it goes from $\tau_{amp}\sim0.6\,\rm \tau_D$ to $\tau_{amp}\sim0.8\,\rm \tau_D$. After a few diffusion times the $\alpha$-quenching stops the amplification, leading the system towards saturation.

A typical example of the spatial distribution behaviour of the magnetic field is reported in Fig. \ref{Bvz_Aq}, where we show the behaviour of the toroidal field strength, in terms of $B_{\rm eq}$, as a function of the $z$ coordinate at different time steps for the configuration Aq. As expected, the field is mainly localized in the NFI regions where $\alpha\neq 0$.

\begin{figure}[!ht]
\begin{center}
\resizebox{\hsize}{!}{\includegraphics{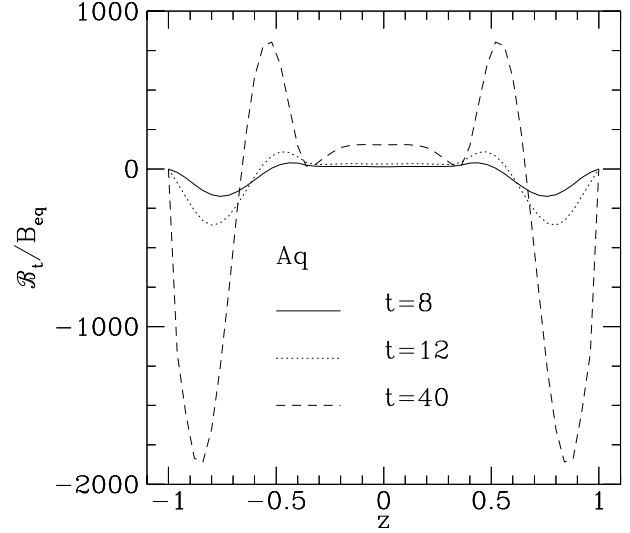}}
\caption{Spatial distribution of the toroidal component of the magnetic field at different time steps, for configuration Aq, with $C_\alpha=10$ and $|q|=0.1$. For this configuration $\alpha$ is different from zero only in regions with $|z|>0.3$.}
\label{Bvz_Aq}
\end{center}
\end{figure}

A couple of remarks are worth making at this point. Firstly, while the figures in this section refer only to the configurations A and Aq, the same behaviour has also been found for the configurations B and Bq, thus indicating that the results presented here are independent of the extent of the NFI zone. Secondly, the extreme magnetic-field amplifications obtained are a direct consequence of the idealized setup used in our modeling of the PNS. We expect that a more realistic description of the geometry of the star and a consistent treatment of the feedback of the magnetic field on the dynamics of the plasma will lower these estimates, yielding magnetic fields which are less strong but still comparable or larger than the equipartition value.

\begin{figure}[!ht]
\begin{center}
\resizebox{\hsize}{!}{\includegraphics{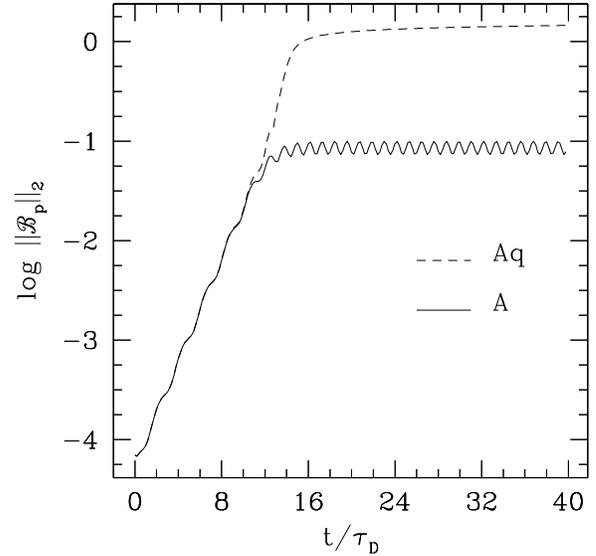}}
\caption{Behaviour of the 2-norm of ${\cal B}_p$ as a function of the number of diffusion times ($t/\tau$) for the A and Aq configurations, $|q| = 2$, $C_\alpha=4$ ($250\,\rm ms \leq P \leq 750\,\rm ms$), and for seed magnetic field strengths of $B_{\rm s}=10^{-4}B_{\rm eq}$. The growth-rate of the magnetic field during its first amplification phase is the same for both configurations, which are eventually driven to saturation by the $\alpha$-quenching .}
\label{norm_dA_AAq}
\end{center}
\end{figure}

\subsection{Asymptotic states of the dynamical configurations \label{Dynamic}}
Although equally idealized, the two {\em dynamical} configurations AB and ABq are expected to provide a better modeling of the first 40 s of life of the PNS, during which the thickness of the NFI region is assumed to vary from about 70\% of the PNS radius, down to about to 20\% of it (\textit{cf.} the profiles of $\alpha$ and $\eta$ in Fig. \ref{aeprofiles}). However, besides the fact that for these configurations it is not possible to define a unique critical period, since it depends on the NFI zone thickness (\textit{cf.} Sect. \ref{TEandCP}), we have found that the qualitative behaviour of the magnetic field for dynamical configurations is very similar to that discussed in the previous section for the static configurations.

More specifically the most salient difference is that, because of the shrinking of the instability zone, the overall amplification is reduced with respect to the A configurations and increased with respect to the B configurations. Indeed, at the beginning of the evolution the NFI zone is as large as that of the A configurations, and at the end it is as large as that of the B configurations. Nevertheless, if one considers a spin rate and a differential rotation strength high enough to have the dynamo mechanism active during all the instability period (\textit {i.e.} $C_\alpha$ and $|q|$ larger than the critical values of the B configurations), then the final field is comparable to or even larger than, that in the static A configurations.

\subsection {The effect of Lorentz force backreaction on differential rotation \label{Lorentz}} 

Several authors (\textit{e.g.},~Weiss et al. \cite{weiss84}; Belvedere
et al. \cite{belv90}; Roald \& Thomas \cite{roald97}; Moss \& Brooke
\cite{moss00}; Gilman \& Rempel \cite{gilman05}; Covas et
al. \cite{covas05}; Rempel \cite{rempel06}) investigated the effect of
the backreaction generated by the Lorentz-force on plasma motion; this
is sometimes referred to as the ``Malkus-Proctor effect''. These
investigations indicate that the dynamo-intensified toroidal field
interacts, via the Lorentz force, with the zonal flow (predominantly azimuthal), thus limiting
the growth of differential rotation and reducing its strength. An
obvious consequence of this effect is that the amplification
of the toroidal field itself is also diminished. All of the above mentioned
studies, which were mainly devoted to the analysis of the Sun or
solar-type stars, made use of an additional Euler equation in
order to include the Lorentz force consistently.

This is different from what is done in our simple, idealized model case,
where we use the induction equation only, and assume the rotational
stress $\Delta \Omega$ to remain constant in time for each
configuration. As a result, any phenomenological, parameterized effect
of the magnetic field on the already fixed $\Delta \Omega$ would not
be consistent with our basic assumptions.

Nevertheless we performed some runs with a phenomenological quenching
function applied to $C_\Omega$ ($\Omega$-quenching), of the same type
as those applied to $\alpha$ and $\eta$. The result of these runs
shows a moderate reduction of the final strength of the toroidal field
for all the examined cases. This result, however, might be somewhat
misleading. Our dynamo is essentially an $\alpha^2\Omega$
one, which can shift to an $\alpha\Omega$ dynamo at strong rotational
stresses, with dynamo numbers (see Sect. \ref{Bfin}) given by the products $C^2_\alpha
C_\Omega$ and $C_\alpha C_\Omega$ respectively. Therefore, the use of
an $\Omega$-quenching applied to $C_\Omega$ essentially amounts to
applying a stronger quenching to $\alpha$ (and this explains the
reduction of the toroidal field strength), but does not limit the fixed
rotational stress.

A rough estimate of the importance of magnetic field backreaction on
differential rotation can be made in terms of the ``Elsasser'' number
$\Lambda = B^2/8\pi\varrho\Omega\nu$ (R\"udiger \& Hollerbach
\cite{RH04}), namely the ratio of magnetic to zonal flow energy. If
$\Lambda \ll 1$ the backreaction effect is negligible, while for
$\Lambda \gg 1$ it is of overwhelming importance, and should produce a
strong reduction of differential rotation. If we define an
equipartition $\Lambda_{\rm eq} = B^2_{\rm
eq}/8\pi\varrho\Omega\eta_0$ and assume that the turbulent viscosity
$\nu$ is of the same order of turbulent magnetic diffusion
(i.e., both momentum and magnetic field are transported by
the same eddies), we have $\Lambda = h^2\Lambda_{\rm eq}$, where $h =
B_t/B_{\rm eq}$. For the typical values of our PNSs given in Sect. \ref{MODEL}, $\Lambda \simeq 0.05$ at equipartition, while for magnetic field strengths exceeding the equipartition value by two or three orders of magnitude, the effect of the Lorentz force on differential rotation cannot be neglected.

Our simplified model does not allow us to predict the intensity of
this effect, even though we obviously expect that the final magnetic
field will be comparable to or slightly larger than the equipartion
one. It is also worth pointing out that a reduction or suppression of
differential rotation does not necessarily imply a suppression of the
dynamo, since the $\alpha^2\Omega$ can shift to a pure $\alpha^2$
dynamo. Furthermore we cannot either predict whether the strong toroidal fields
generated are stable against the magneto-rotational or Tayler
instabilities, which would require a dynamical analysis, beyond the
scope of the present exploratory study; this will be addressed in a more realistic model we intend to develop in the future.

\subsection{A general expression for the final magnetic field \label{Bfin}}
From the discussion about the time evolution of the solutions of
Eqs.~(\ref{EQFINA}) and (\ref{EQFINB}) (\textit{cf.} Sect.
\ref{TEandCP}) and about their asymptotic states (\textit{cf.}
Sect. \ref{Static} and \ref{Dynamic}), it is apparent that the
configurations considered in this work have common features that we
believe reflect a fundamental behaviour of the mean field dynamo
process. Thus we expect these features to be present also when the toy
model considered here is replaced by a more realistic one. In
what follows we discuss how to summarize these analogies by presenting
a general expression for the final magnetic field.

Using Figs.~\ref{A_tor} and ~\ref{A_pol} as guides, it is easy to recognize the existence of a transition value $|q^*|$ such that for
$|q| \lesssim |q^*|$ the final magnetic field does not depend on the degree of differential rotation, while for $|q| \gtrsim |q^*|$ it increases as a power-law. We therefore express this increase with a phenomenological relation of the type
\begin{eqnarray}
\label{phen_1}
&&{\cal B}_t^{\rm fin} = K_{t}\; (C_\alpha)^{\delta_t} |q|^{\gamma_t} \;,
\\
\label{phen_2}
&&{\cal B}_p^{\rm fin} = K_{p}\; (C_\alpha)^{\delta_p} |q|^{\gamma_p} \;,
\end{eqnarray}
where the indices $t$ and $p$ refer to the toroidal and poloidal components, respectively. Note that the threshold value $|q^*|$ may
well differ between (\ref{phen_1}) and (\ref{phen_2}), and that the constants $K_{t,p}$ depend on the particular configuration, \textit {i.e.} on the profile of $\alpha$ and $\eta$ and on $\eta$-quenching.

\begin{table}[h]
\caption{Indices of the power-law behaviour found in the dependence of $B^{\rm fin}$ on the degree of differential rotation $|q|$ and the spin rate $C_\alpha$. The errors, not reported in the table, are of a few percent only.}
\begin{center}
\begin{tabular}{ccccc}
\hline
\hline
  & $\gamma_t$  & $\gamma_p$  & $\delta_t$ & $\delta_p$ \\
\hline
A   & $0.53 $ & $-0.48 $ & 1.00 & 0.03 \\
B   & $0.51 $ & $-0.49 $ & 1.00 & 0.00 \\
AB  & $0.49 $ & $-0.48 $ & 0.94 & 0.09 \\
\hline
Aq  & $0.77 $ & $-0.37 $ & 1.10 & 0.06 \\
Bq  & $0.54 $ & $-0.48 $ & 1.09 & 0.07 \\
ABq & $0.56 $ & $-0.46 $ & 1.13 & 0.12 \\
\hline
\end{tabular}
\end{center}
\label{slopes}
\end{table}

Using a power-law fit, we have calculated the values for the exponents in the phenomenological expressions (\ref{phen_1}) and (\ref{phen_2}) and collected them in Table~\ref{slopes} for both the static and the dynamical models, with and without $\eta$-quenching. The reported values of $\gamma_{t,p}$ have been computed using several configurations having different $C_{\alpha}$ (between $6$ and $200$), while those for $\delta_{t,p}$ have been derived from configurations differing by the amount of differential rotation ($|q|$ in the range $10^{-1}-10^2$). For all of the configurations considered, the variance around the reported values is very small and of a few percent only.

Overall, the data in Table~\ref{slopes} show that both $\gamma_t$ and $\gamma_p$ are very close to either $1/2$ or $-1/2$ (with the exception of Aq, for which they are $\sim 0.75$ and $\sim - 0.35$, respectively) and that quite generically $|\gamma_t| + |\gamma_p| \simeq 1$. We can therefore rewrite expressions (\ref{phen_1})--(\ref{phen_2}) simply as 
\begin{eqnarray}
\label{phen_3}
&&{\cal B}_t^{\rm fin} \simeq K_{t}\; C_\alpha |q|^{1/2} \;,
\\
\label{phen_4}
&&{\cal B}_p^{\rm fin} \simeq K_{p}\; |q|^{-1/2} \;.
\end{eqnarray}

The behaviour of the toroidal and poloidal final fields in the
super-critical dynamo regime, as given by Eqs. (\ref{phen_3}) and
(\ref{phen_4}), can be understood in terms of the characteristic
dynamo parameters, $C_\alpha$ and $C_\Omega$. The first parameter represents the
dynamo's ability to regenerate the poloidal field from the toroidal
one (i.e., $\alpha$-effect), while the second parameter the
dynamo's ability to regenerate the toroidal magnetic field from its
parent poloidal field, via the action of differential rotation
(i.e., $\Omega$-effect). Both parameters are important as
they define the \textit{``dynamo number''}, given by $N_{\alpha\Omega}
\equiv C_\alpha C_\Omega$ for an $\alpha\Omega$ dynamo, and
$N_{\alpha^2\Omega}=C^2_\alpha C_\Omega$ for an $\alpha^2\Omega$ dynamo. In essence, below a certain critical number $N_{\rm c}$, which depends on the specific model considered, no dynamo action is possible,
while the dynamo becomes ever more efficient as $N$ increases.

In our case, when the effect of differential rotation becomes
important (i.e., high $|q|$ values), our dynamo behaves
essentially as an $\alpha\Omega$ dynamo with $C_\Omega/C_\alpha \simeq
5\,|q| \gg 1$ [see Eq. (\ref{qfactor})]. For this dynamo, the ratio of
the toroidal to the poloidal field, ${\cal B}_t/{\cal B}_p \simeq
(C_\Omega/C_\alpha)^{1/2}$, is high, as verified also with our
nonlinear calculations (see Figs. \ref{A_tor} and \ref{A_pol}), with
typical values ranging from 20 to 30.  The relationships described by Eqs. (\ref{phen_3}) and (\ref{phen_4}) have been derived in the
super-critical regime region where ${\cal B}_t$ substantially
increases and ${\cal B}_p$ decreases with increasing $C_\Omega$ or
$|q|$ for each fixed $C_\alpha$ value (see Figs. \ref{A_tor} and
\ref{A_pol} for $\log |q| > 0$). Therefore, it is not surprising that
${\cal B}_t \propto N^{1/2}$ [Eq. (\ref{phen_3})], namely that the
toroidal field increases with dynamo efficiency. On the other hand,
the poloidal field ${\cal B}_p \propto C_\alpha/N^{1/2}$
[Eq. (\ref{phen_4})] will necessarily decrease with dynamo efficiency
increase, since the increase of $N$ is only due to the increase of
$C_\Omega$ ($C_\alpha$ is kept fixed for each simulation).  This is
consistent with the fact that, at high differential rotation rates,
the zonal flow dominates with respect to the vertical motions (predominantly radial) thus
reducing the strength of convection and therefore the efficiency of
the $\alpha$-effect, namely the regeneration of the poloidal
field.

We expect that more sophisticated calculations will produce changes of
these values reported in Eqs. (\ref{phen_3}) and (\ref{phen_4}),
especially in the transition between the $\alpha^2\Omega$ and the
$\alpha\Omega$ dynamos, and when the Lorentz-force feedback is
properly taken into account. However, we expect that the behaviour of
the magnetic field as given by Eqs. (\ref{phen_3}) and (\ref{phen_4})
will remain unaltered.

\section{Conclusions \label{conclusion}}

We have presented a toy model to describe the amplification of the magnetic field inside a proto-neutron star (PNS) via a dynamo action. The model assumes that a neutron-finger instability (NFI) develops in the outer regions of a PNS during the early stages of its life as discussed by Miralles et al.~(\cite{miralles00}, \cite{miralles02}), and that the conditions for the generation of a mean-field dynamo process are met.  Although highly simplified by being only one-dimensional and by adopting the kinematic approximation, our model aims to capture the qualitative features of the dynamo action by including a moving boundary of the instability zone and the nonlinearities introduced by the feedback processes, which saturate the growth of the magnetic field (i.e. $\alpha$-quenching) and suppress its turbulent diffusion (i.e. $\eta$-quenching).

In essence, the amplification of the magnetic field is described in terms of a system of coupled partial differential equations of mixed hyperbolic-parabolic type, which are solved numerically for a very large variety of initial conditions. These include varying the spin period of the PNS, the strength of the differential rotation between the core and the surface, the intensity of the primordial (seed) magnetic field, and the extent of the NFI zone.

Overall, we have found that, independently of whether the size of the NFI zone varies in time or not, the amplification of the magnetic field undergoes a first exponential increase with growth-time that is the same for both the toroidal and the poloidal components of the magnetic field. The exact value of the growth-time depends on several parameters and it is roughly in the range $\tau_{amp} \sim \left[ 0.5-2.5 \right]\, \tau_D$, with $\tau_D$ being the diffusion timescale. The exponential growth then stops through the back reaction of the $\alpha$-quenching and the magnetic field reaches saturation. The final magnetic field produced at the end of the 40 s of evolution does not depend sensitively on the initial magnetic field, but it does depend on whether the $\eta$-quenching is active or not, becoming 2-3 orders of magnitude larger in the first case.

Despite its crudeness, our model is also able to capture another important feature of the dynamo mechanism, namely the existence of a
critical rotation period $P_c$, above which no dynamo action is possible and the magnetic field simply decays (Bonanno et al. \cite{bonanno03}). For periods near the critical one, on the other hand, the dynamo is just able to sustain the magnetic field close to its initial value, thus avoiding its decay. However, as the spin rate (or the degree of differential rotation) is increased, the dynamo becomes more and more efficient, amplifying the magnetic field up to values several orders of magnitude larger than the equipartion magnetic field. These very high intensities ($10^{18}\,G$ for the toroidal component and $10^{14}\,G$ for the poloidal one) may seem unphysical at first sight; however, here we are considering magnetic fields still inside the neutron star and only the poloidal component is thought to emerge afterwards. Determining the critical period accurately is important to constrain the fraction of neutron stars that may undergo this magnetic-field amplification at birth, and we have found that $P_c$ is in the range $33-600 \,\rm ms$ for rigidly rotating PNSs, becoming larger as the degree of differential rotation is increased.  As a result, as long as we have no general constraints on the strength of the differential rotation, a lower limit of $P \simeq 30$ ms can be taken as the generic threshold below which a mean-field dynamo may be active in a newly born PNS.

Another interesting result of this investigation is that, despite the large parameter space considered, the final value of the magnetic field seems to follow a surprisingly robust dependence with the spin period and the degree of differential rotation, both of which can be summarized in a phenomenological expression of the type
\begin{equation}
\label{scaling}
{\cal B}^{fin} \propto (C_\alpha)^\delta |q|^\gamma
\end{equation}
which holds only for $|q|$ larger than a transition level of differential rotation $|q^*|$, whose exact value depends on the configuration and on the other parameters (overall it is in the range $10^{-2}-10^1$). The exponents $\delta,\ \gamma$ are different for the toroidal and poloidal magnetic field components and depend only very weakly on all of the parameters varied in this analysis. In particular, for the toroidal magnetic-field component we have found $\delta \sim 1$ and $\gamma\sim 1/2$, while for the poloidal one $\delta \sim 0$ and $\gamma\sim -1/2$. The exact values of these exponents are likely to be modified by more realistic and multidimensional calculations, but we also expect that the scaling in expression (\ref{scaling}) will persist in further refinements of this treatment.

The work presented here can be improved in a number of different ways. A first possibility is a more realistic description of the geometry of the problem, with a two or three-dimensional description of the PNS. A second and computationally less expensive alternative is that of improving the nonlinear feedback of the magnetic field on the dynamics of the matter. This can be done by using the same geometry adopted here, but coupling the mean-field induction equation (\ref{induction}) with the solution of the MHD equations for the conservation of energy and momentum.

\begin{acknowledgements}
We are grateful to V. Urpin and J. C. Miller for their useful comments
and suggestions. This work was partially supported by the Italian
Ministry of University and Research under the contract PRIN
2004024993.

\end{acknowledgements}

\bibliographystyle{aa}

\end{document}